\documentclass[12pt]{iopart}

\usepackage{iopams,graphicx}  
\begin{document}

\title[Doping-dependent magnetization plateaus]{Doping-dependent magnetization plateaus of a coupled spin-electron chain: exact results}

\author{Jozef Stre\v{c}ka, Jana \v{C}is\'{a}rov\'{a}}
\address{Department of Theoretical Physics and Astrophysics, Faculty of Science, P. J. \v{S}af\'{a}rik University, Park Angelinum 9, 040 01 Ko\v{s}ice, Slovakia}
\ead{jozef.strecka@upjs.sk}
\vspace{10pt}

\begin{abstract}
A coupled spin-electron chain composed of localized Ising spins and mobile electrons is exactly solved in an external magnetic field within the transfer-matrix method. The ground-state phase diagram involves in total seven different ground states, which differ in the number of mobile electrons per unit cell and the respective spin arrangements. A rigorous analysis of the low-temperature magnetization process reveals doping-dependent magnetization plateaus, which may be tuned through the density of mobile electrons. It is demonstrated that the  fractional value of the electron density is responsible for an enhanced magnetocaloric effect due to an annealed bond disorder of the mobile electrons.   
\end{abstract}

\pacs{75.10.Pq, 75.30.Sg, 75.40.Cx, 75.10.Jm}
\vspace{2pc}
\noindent{\it Keywords}: coupled spin-electron chain, localized spins, mobile electrons, magnetization plateaus, doping 
\vspace{1pc}												
\submitto{\MRE}
%
%
%
\section{Introduction}

The magnetization process of strongly correlated electron systems has received a lot of attention, because it may involve at low enough temperatures intermediate plateaus  as a profound manifestation of peculiar quantum ground states \cite{scho04,hone04,tsve01,haip15}. An existence of the intermediate magnetization plateaus, which refer to regions with a constant magnetization spread over a finite range of the magnetic fields, has been convincingly evidenced in a variety of one-dimensional quantum systems including spin chains, ladders and other related systems \cite{hida94,tots97,cabr98,saka98,momo99}. It is well established that the magnetization plateaus of quantum spin chains do occur just at rational fractions of the saturation magnetization, whereas their position is subject to a quantization condition \cite{oshi97} derived by generalizing Lieb-Schultz-Mattis theorem \cite{lieb61}. 

A special mechanism leading to intermediate magnetization plateaus controlled by doping has been studied within integrable spin-$S$ chains doped with spin-1/2 mobile carriers \cite{frah99,lama11}. It has been evidenced that the doping within a relatively large range of carrier concentrations leads to the magnetization plateaus at a tunable fraction of the saturation magnetization in the high-field region \cite{frah99,lama11}. However, the doping may also allow a continuous variation of the magnetization plateaus even in the low-field region as exemplified by rigorous analytical and numerical calculations for the modulated Hubbard chains \cite{cabr00,cabr01} and ladders \cite{cabr02,roux06,roux07}. 

In the present work we will propose and exactly solve a correlated spin-electron chain, which involves the localized Ising spins situated at nodal lattice sites and mobile electrons delocalized over the pairs of decorating sites placed at each its bond. Our main goal is to examine the low-temperature magnetization process of this system as a function of the electron density with the aim to bring insight into doping-dependent magnetization plateaus. 

The outline of this paper is as follows. In Section \ref{sec:model}, the model under investigation will be defined and the method used for its exact solution will be briefly described. The ground-state phase diagram and respective spin arrangement realized within particular ground states will be discussed in Section \ref{sec:results} along with the magnetization process and magnetocaloric effect. Finally, the most significant findings will be summarized in Section \ref{sec:conclusions}.

\section{Model and method}
\label{sec:model}

\begin{figure}
\begin{center}
\includegraphics[width=10.0cm]{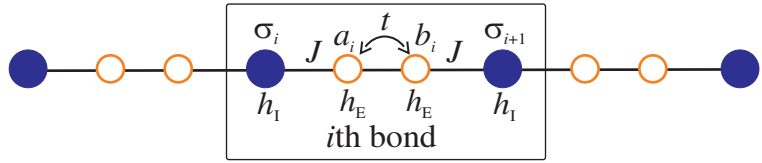}
\end{center}
\vspace{-0.4cm}
\caption{\small (Color online) A schematic representation of the studied spin-electron linear chain. The large filled circles denote nodal lattice sites occupied by the localized Ising spins, while the small open circles denote the decorating sites over which the mobile electrons are delocalized. A part of the system demarked by the rectangle is described by the bond Hamiltonian (\ref{Hi}).}
\label{fig1}
\end{figure}

Let us define a coupled spin-electron chain in an external magnetic field, which involves localized Ising spins situated on its nodal sites and mobile electrons delocalized over the pairs of decorating sites as schematically shown in Fig.~\ref{fig1}. 
For further convenience, the total Hamiltonian of the investigated spin-electron model can be written as a sum over the bond Hamiltonians  
\begin{eqnarray}
\hat{\cal H} = \sum_{i=1}^N \hat{\cal H}_i,
\label{htot}
\end{eqnarray}
whereas each bond Hamiltonian $\hat{\cal H}_i$ contains all interaction terms involving the mobile electrons from the $i$-th bond (see Fig.~\ref{fig1})
\begin{eqnarray}
\hat{\cal H}_i = &-& t ( \hat{a}_{i,\uparrow}^{\dagger} \hat{b}_{i,\uparrow} + \hat{a}_{i,\downarrow}^{\dagger} \hat{b}_{i,\downarrow} 
+ \hat{b}_{i,\uparrow}^{\dagger} \hat{a}_{i,\uparrow} + \hat{b}_{i,\downarrow}^{\dagger} \hat{a}_{i,\downarrow} ) \nonumber \\
&-& J [ \sigma_i^z (\hat{n}_{ai,\uparrow} - \hat{n}_{ai,\downarrow}) + \sigma_{i+1}^z (\hat{n}_{bi,\uparrow} - \hat{n}_{bi,\downarrow}) ]/2 \nonumber \\
&-& h_{\rm E} [ (\hat{n}_{ai,\uparrow} - \hat{n}_{ai,\downarrow}) + (\hat{n}_{bi,\uparrow} - \hat{n}_{bi,\downarrow})]/2 
 - h_{\rm I} (\sigma_i^z + \sigma_{i+1}^z)/2.
\label{Hi}
\end{eqnarray}
Here, the symbols $\hat{\alpha}_{i,\gamma}^{\dagger}$ and $\hat{\alpha}_{i,\gamma}$ denote standard creation and annihilation fermionic operators for the mobile electrons occupying the decorating sites $\alpha = \{ a,b \}$ with the spin orientation $\gamma = \{ \uparrow,\downarrow \}$ and $\hat{n}_{\alpha i,\gamma} = \hat{\alpha}_{i,\gamma}^{\dagger} \hat{\alpha}_{i,\gamma}$ represents the respective number operator with eigenvalues $n_{\alpha i, \gamma} = \{ 0,1\}$. The symbol $\sigma_i^z = \pm 1/2$ corresponds to the Ising spin situated at $i$-th nodal lattice site. The parameter $t$ thus represents the hopping term accounting for the kinetic energy of the mobile electrons and the parameter $J$ describes the Ising-type interaction between the nearest-neighbour localized spins and mobile electrons. For the purpose of the future calculations, we have distinguished between the magnetic field $h_{\rm I}$ acting on the localized Ising spins and the magnetic field $h_{\rm E}$ acting on the mobile electrons though both magnetic fields will be finally set equal to each other $h_{\rm I}=h_{\rm E}=h$ at the end of the calculation. 

The crucial step in finding of the exact solution of the correlated spin-electron chain is to calculate the grand-canonical partition function  
\begin{eqnarray}
\Xi = \sum_{\{ \sigma_i \}} {\rm Tr} \exp \left(-\beta \hat{\cal H} + \beta \mu \sum_{i=1}^N {\hat n}_i \right),
\label{gcpf}
\end{eqnarray}
where $\beta = 1/(k_{\rm B} T)$, $k_{\rm B}$ is Boltzmann's constant, $T$ stands for the absolute temperature, $\sum_{\{ \sigma_i \}}$ denotes a summation over all possible configurations of the Ising spins and the symbol ${\rm Tr}$ stands for a trace over degrees of freedom of all mobile electrons. Furthermore, the number operator ${\hat{n}}_i = \sum_{\gamma = \{ \uparrow, \downarrow \}} (\hat{n}_{a i,\gamma} + \hat{n}_{b i,\gamma})$ corresponds to the total number of the mobile electrons on $i$-th bond and $\mu$ is the respective chemical potential, which allows to tune the electron density. Due to the commuting character of different bond Hamiltonians $[{\hat{\cal H}}_i,{\hat{\cal H}}_j] = 0$ ($i \neq j$), the grand-canonical partition function $\Xi$ can be partially factorized into the product
\begin{eqnarray}
\Xi = \sum_{\{ \sigma_i \}} \prod_{i=1}^N {\rm Tr}_i \exp(-\beta \hat{\cal H}_i) \exp(\beta \mu {\hat n}_i),
\label{gcpf2}
\end{eqnarray}
where the symbol ${\rm Tr}_i$ is used to denote a partial trace over degrees of freedom of the mobile electrons from the $i$-th bond. An explicit form of the grand-canonical partition function $\Xi$ can be found with the use of the standard transfer-matrix approach \cite{baxt82}. All subsequent steps of the further calculation are quite analogous to the ones reported in our preceding work for the zero-field case \cite{cisa14} and hence, we will only discuss them briefly here. The bond Hamiltonian $\hat{\cal H}_i$ commutes with the $z$-component of the total spin operator ${\hat{S}}_i^z = \sum_{\alpha = \{ a,b \}} ({\hat{n}}_{\alpha i, \uparrow} - {\hat{n}}_{\alpha i, \downarrow})/2$ of the mobile electrons from the $i$-th bond, i.e. $[\hat{\cal H}_i, {\hat S}_i^z] = 0$. Owing to this fact, the eigenvalues of the bond Hamiltonian $\hat{\cal H}_i$ for the coupled spin-electron chain in an external magnetic field can be simply obtained by shifting the respective eigenvalues of the zero-field model \cite{cisa14} by the appropriate Zeeman's terms. Using this procedure, one obtains the following energy spectrum  
\begin{eqnarray}
\fl E_{i1} = E_{i8} = E_{i9} = E_{i16} = - \frac{h_{\rm I}}{2} (\sigma_i^z + \sigma_{i+1}^z), \nonumber \\
\fl E_{i2,i3} = E_{i12,i13} = - \frac{h_{\rm I}}{2} (\sigma_i^z + \sigma_{i+1}^z) - \frac{1}{2} \left [h_{\rm E} + (h_i + h_{i+1}) \pm \sqrt{(h_i - h_{i+1})^2 + 4t^2} \right], \nonumber \\
\fl E_{i4,i5} = E_{i14,i15} = - \frac{h_{\rm I}}{2} (\sigma_i^z + \sigma_{i+1}^z) + \frac{1}{2} \left [ h_{\rm E} + (h_i + h_{i+1}) \pm \sqrt{(h_i - h_{i+1})^2 + 4t^2} \right ], \nonumber \\
\fl E_{i6,i7} = - \frac{h_{\rm I}}{2} (\sigma_i^z + \sigma_{i+1}^z) \pm [h_{\rm E} + (h_i + h_{i+1})], \nonumber \\
\fl E_{i10,i11} = - \frac{h_{\rm I}}{2} (\sigma_i^z + \sigma_{i+1}^z) \pm \frac{1}{2} \sqrt{(h_i - h_{i+1})^2 + 4t^2}.
\label{ES} 
\end{eqnarray}
Here, we have introduced the following notation $h_i = J \sigma_i^z / 2$ in order to write the relevant eigenvalues in a more abbreviated form, whereas the notation for lower indices follows from Ref. \cite{cisa14} (it determines the sectors with different number of the mobile electrons per unit cell). At this stage, the eigenvalues (\ref{ES}) of the bond Hamiltonian (\ref{Hi}) can be employed for a calculation of the expression entering the right-hand-side of the factorized form (\ref{gcpf}) of the grand-canonical partition function 
\begin{eqnarray}
\fl T(\sigma_i^z,\sigma_{i+1}^z) &=& {\rm Tr}_i \exp(-\beta \hat{\cal H}_i) \exp(\beta \mu \hat{n}_i) = \sum_{j=1}^{16} \exp(-\beta E_{ij}) z^{n_{ij}}
\nonumber \\ 
\fl &=& \exp \left[\frac{\beta h_{\rm I}}{2} \left(\sigma_i^z + \sigma_{i+1}^z\right) \right] 
\Bigl\{ 1 + 2z^2 + z^4 + 2 z^2 \cosh [\beta (h_E + h_i + h_{i+1})] \Bigr. \nonumber \\ 
\fl &+& 4 \left(z + z^3\right) \cosh \left[\frac{\beta}{2} \sqrt{(h_i - h_{i+1})^2 + 4t^2}\right] 
\cosh \Bigl[\beta \left(h_E + h_i + h_{i+1}\right) \Bigr] \nonumber \\
\fl &+& \left. 2 z^2 \cosh \left[ \frac{\beta}{2}\sqrt{(h_i - h_{i+1})^2 + 4t^2}\right] \right\}.
\label{tmd}
\end{eqnarray}
The parameter $z = \exp (\beta \mu)$ denotes the fugacity of the mobile electrons. After substituting Eq. (\ref{tmd}) into Eq. (\ref{gcpf2}) and performing successive summation over spin degrees of freedom of individual Ising spins one obtains 
\begin{eqnarray}
\Xi = \sum_{\{ \sigma_i \}} \prod_{i=1}^{N} T(\sigma_i^z,\sigma_{i+1}^z) = \mbox{Tr} \, T^N,
\label{gcpf3}
\end{eqnarray}
since the expression  $T(\sigma_i^z,\sigma_{i+1}^z)$ can be identified as the usual transfer matrix \cite{baxt82} 
\begin{eqnarray}
T(\sigma_i^z,\sigma_{i+1}^z) = 
\left(
  \begin{array}{cc}
     {\rm T}(1/2,1/2) & {\rm T}(1/2,-1/2) \\     
		 {\rm T}(-1/2,1/2) & {\rm T}(-1/2,-1/2)
  \end{array}
\right) = 
\left(
  \begin{array}{cc}
     V_1 & V_3 \\     
		 V_3 & V_2
  \end{array}
\right).
\label{tm}
\end{eqnarray}
The transfer-matrix elements are explicitely given by
\begin{eqnarray}
V_1 &=& \exp \left(\frac{\beta h_{\rm I}}{2}\right) \left\{ 1 + 2z^2 + z^4 
+ 4 (z + z^3) \cosh \left[\frac{\beta}{4} (2 h_{\rm E} + J)\right] \cosh(\beta t) \right. \nonumber \\
&+& \left. 2 z^2 \cosh \left[\frac{\beta}{2} (2 h_{\rm E} + J)\right] + 2 z^2 \cosh(2 \beta t)  \right\}, \nonumber \\
V_2 &=& \exp \left(-\frac{\beta h_{\rm I}}{2}\right) \left\{ 1 + 2z^2 + z^4 + 4 (z + z^3) \cosh \left[\frac{\beta}{4} (2 h_{\rm E} - J)\right] \cosh(\beta t) \right. \nonumber \\
&+& \left. 2 z^2 \cosh \left[\frac{\beta}{2} (2 h_{\rm E} - J)\right] + 2 z^2 \cosh(2 \beta t) \right \}, \nonumber \\
V_3 &=& 1 + 2 z^2 + z^4 + 4 (z + z^3) \cosh(\beta h_{\rm E}/2) \cosh(\beta P/4) \nonumber \\
&+& 2 z^2 \cosh(\beta h_{\rm E}) + 2 z^2 \cosh(\beta P/2),
\label{V123}
\end{eqnarray}
where the parameter $P = \sqrt{J^2 + (4 t)^2}$. In the thermodynamic limit ($N \to \infty$), the grand potential (per elementary unit cell) associated to the grand-canonical partition function (\ref{gcpf3}) will take a simple form
\begin{eqnarray}
\Omega = - k_{\rm B} T \lim_{N \to \infty} \frac{1}{N} \ln \Xi = -k_{\rm B} T \ln \lambda_{max},
\label{grandpotr}
\end{eqnarray}
whereas the larger eigenvalue $\lambda_{max}$ of the transfer matrix (\ref{tm}) is given by
\begin{eqnarray}
\lambda_{max} = \frac{1}{2} \left[ V_1 + V_2 + \sqrt{{(V_1 - V_2)}^2 + 4 V^2_3} \right].
\label{lmax}
\end{eqnarray}
The exact result (\ref{grandpotr}) for the grand potential can be subsequently employed in order to determine the electron density per one couple of the decorating sites
\begin{eqnarray}
\rho = \langle \hat{n}_i \rangle = - {\left(\frac{\partial \Omega}{\partial \mu} \right)}_{T},
\label{density}
\end{eqnarray}
which reflects the doping of the coupled spin-electron chain with the mobile electrons.

Our particular interest will be successively focused on a magnetization process, so let us calculate separate contributions of the localized Ising spins and the mobile electrons to the total magnetization. The sublattice magnetization $m_{\rm I}$ of the localized Ising spins and the sublattice magnetization $m_{\rm E}$ of the mobile electrons per elementary unit cell can be obtained from the following relations
\begin{eqnarray}
m_{\rm I} = -\left(\frac{\partial \Omega}{\partial h_{\rm I}}\right)_{z}, \qquad m_{\rm E} = -\left(\frac{\partial \Omega}{\partial h_{\rm E}}\right)_{z}.
\label{MIEdef}
\end{eqnarray}  
The final exact expressions for the sublattice magnetizations $m_{\rm I}$ and $m_{\rm E}$ are listed below 
\begin{eqnarray}
m_{{\rm I}} &=& \frac{1}{2} \frac{V_1 - V_2}{\sqrt{(V_1 - V_2)^2 + 4 V^2_3}}, \nonumber \\
m_{{\rm E}} &=& \frac{(V_1 - V_2)(W_1 - W_2) + 4 V_3 W_3 + (W_1 + W_2) \sqrt{(V_1 - V_2)^2 + 4 V^2_3}}
                     {(V_1 - V_2)^2 + 4 V^2_3 + (V_1 + V_2) \sqrt{(V_1 - V_2)^2 + 4 V^2_3}}.  
\label{Mag3D}
\end{eqnarray}
The latter expression involves the new parameters $W_1$, $W_2$ and $W_3$ defined as
\begin{eqnarray}
W_1 &=& \frac{\partial{V_1}}{\partial h_{\rm E}} = \exp (\beta h_{\rm I}/2) Q^{+}, \qquad 
W_2 = \frac{\partial{V_2}}{\partial h_{\rm E}} = \exp (-\beta h_{\rm I}/2) Q^{-}, \nonumber \\
W_3 &=& \frac{\partial{V_3}}{\partial h_{\rm E}} = 2 z (1 + z^2) \sinh (\beta h_{\rm E}/2) \cosh (\beta P/4) 
+ 2 z^2 \sinh (\beta h_{\rm E}), \label{Vp} \\
Q^{\pm} &=& z (2 + z^2/2) \cosh (\beta t) \sinh [\beta (2 h_{\rm E} \pm J)/4]/2 + 2 z^2 \sinh [\beta (2 h_{\rm E} \pm J)/2]. \nonumber 
\end{eqnarray}
It is worth recalling that the sublattice magnetization $m_{\rm E}$ actually gives the overall magnetization of the mobile electrons per one couple of the decorating sites. The total magnetization $m_T$ of the coupled spin-electron chain per unit cell is then given by
\begin{eqnarray}
m_T = m_{\rm I} + m_{\rm E}
\label{MT}
\end{eqnarray} 
and the total magnetization normalized with respect to its saturation value $m_S = (1 + \rho)/2$ follows from
\begin{eqnarray}
\frac{m_T}{m_S} = \frac{2(m_{\rm I} + m_{\rm E})}{1 + \rho}.
\label{MTMS}
\end{eqnarray}
Last but not least, we have exactly calculated the entropy of the studied spin-electron chain according to the basic thermodynamic relation $S = - \frac{\partial \Omega}{\partial T}$ \cite{huan63}, but the final expression is too cumbersome to write it down here explicitly.

\section{Results and discussion}
\label{sec:results}

This section will be devoted to a discussion of the most interesting results obtained for the coupled spin-electron linear chain with the ferromagnetic interaction between the localized Ising spins and mobile electrons, which will hereafter serve as the energy unit (i.e. $J=1$). For the sake of simplicity, the Boltzmann's constant is set to unity (i.e. $k_{\rm B}=1$) and the local magnetic fields acting on the localized Ising spins and mobile electrons are set equal to each other $h = h_{\rm I} = h_{\rm E}$ to avoid overparametrization. 

\subsection{Ground state}

\begin{figure}
\begin{center}
\includegraphics[width=10.0cm]{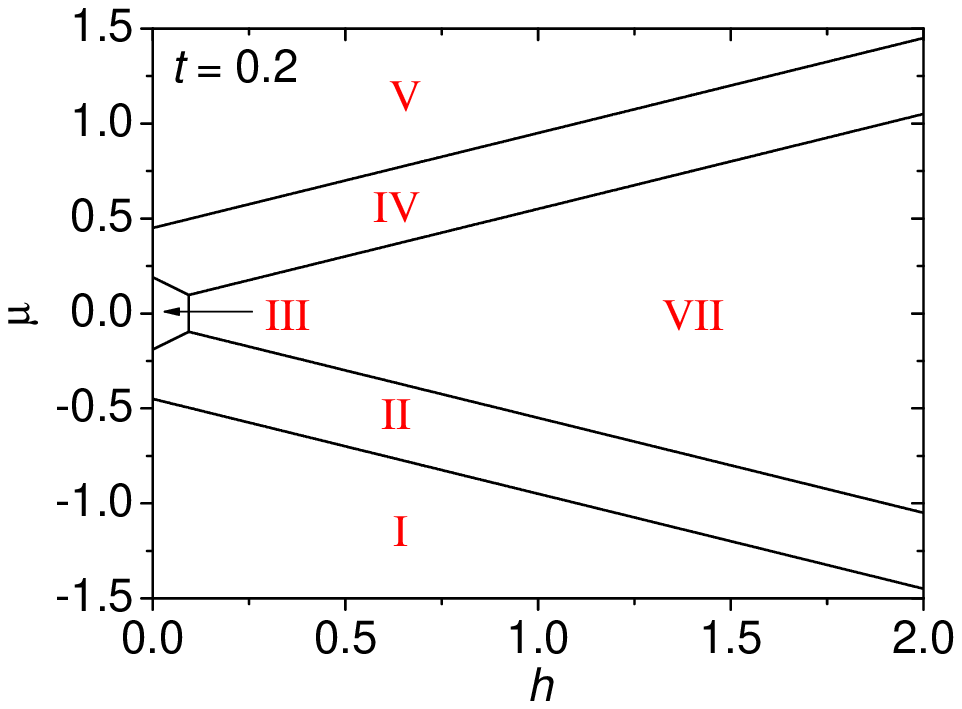}
\vspace{-0.6cm}
\includegraphics[width=10.0cm]{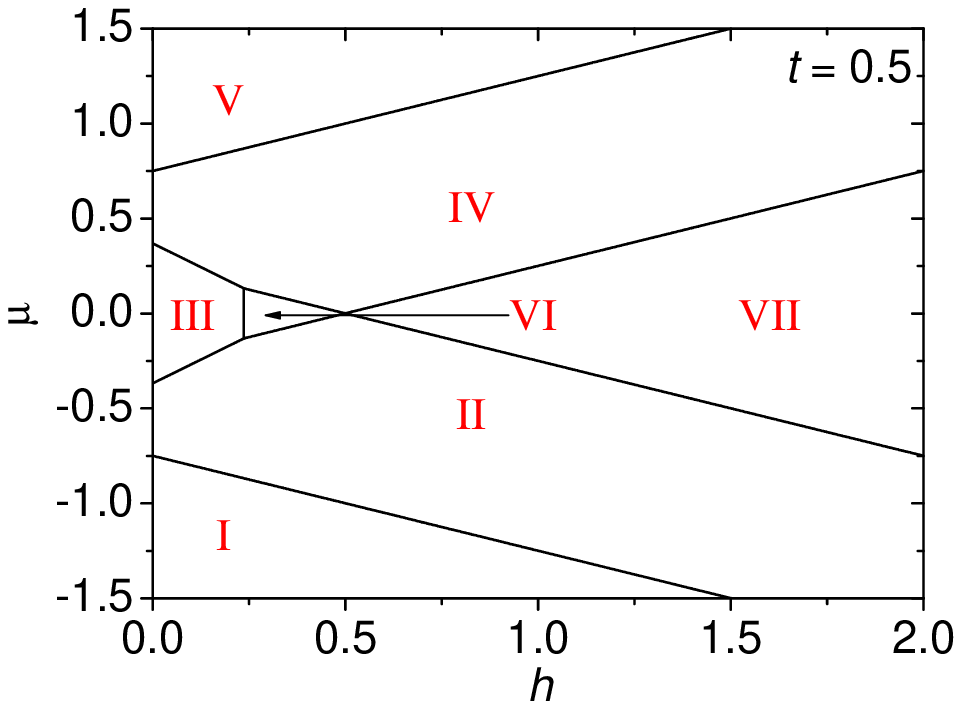}
\vspace{-0.6cm}
\includegraphics[width=10.0cm]{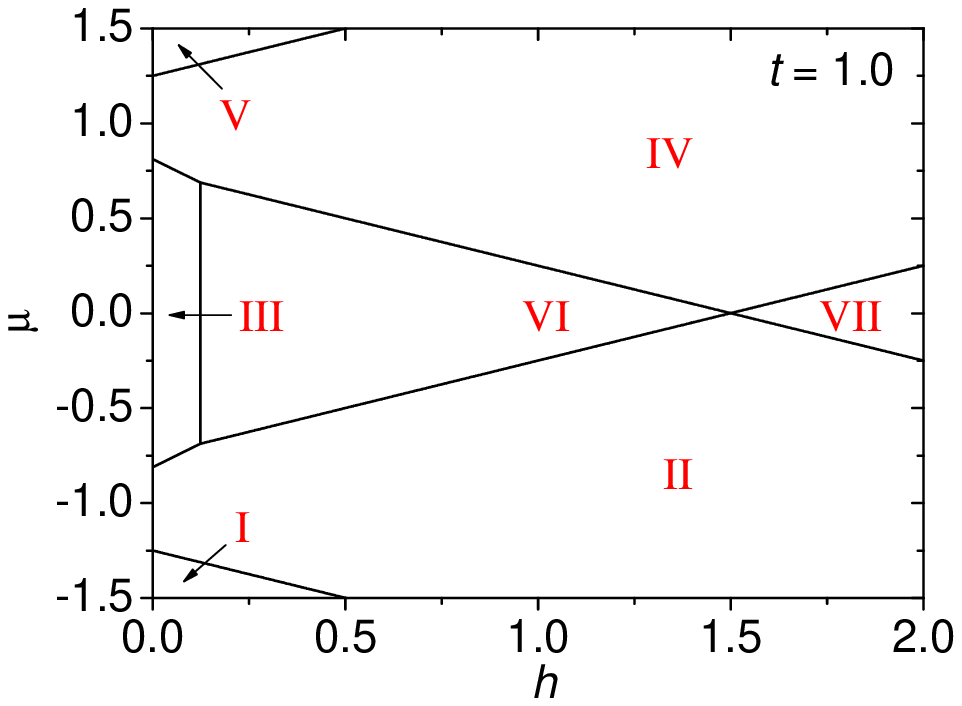}
\end{center}
\vspace{-0.6cm}
\caption{\small (Color online) The ground-state phase diagram of the coupled spin-electron chain in the $h-\mu$ plane for three different values of the hopping term $t = 0.2$, $0.5$ and $1.0$.}
\label{fig2}
\end{figure}
Let us begin with a detailed analysis of the ground-state phase diagram, which is depicted in Fig.~\ref{fig2} in the magnetic field versus chemical potential ($h-\mu$) plane for three different values of the hopping term $t = 0.2, 0.5$ and $1.0$. As one can see, there are seven different ground states of the studied spin-electron system enumerated as $\rm I$ - $\rm VII$. The constructed ground-state phase diagrams are symmetric with respect to $\mu = 0$ axis due to the particle-hole symmetry, which connects the phases ${\rm I}$ and ${\rm V}$, as well as, the phases ${\rm II}$ and ${\rm IV}$. Owing to a commuting character of different bond Hamiltonians, the ground state of the full spin-electron chain can be written as a tensor product over the lowest-energy eigenstate (\ref{ES}) of the bond Hamiltonian (\ref{Hi}). In the following, we will specify eigenvectors of all individual ground states along with their respective eigenenergies per elementary unit cell. 

The phase ${\rm I}$ with the character of disordered paramagnetic phase becomes the ground state for sufficiently low values of the chemical potential
\begin{eqnarray}
{|{\rm I} \rangle} &=& \Biggl \{ \begin{array}{l}
\displaystyle \prod_{i=1}^N {\Bigl|\frac{1}{2}\Bigr\rangle}_{\sigma_i} \otimes {|0,0 \rangle}_i \quad \qquad (\mbox{for} \,\, h > 0), \\
\displaystyle\prod_{i=1}^N {\Bigl|\pm \frac{1}{2}\Bigr\rangle}_{\sigma_i} \otimes {|0,0 \rangle}_i \qquad (\mbox{for} \,\, h = 0), \end{array} \Biggr.
\nonumber \\
E_{\rm I} &=& -\frac{h}{2},
\label{GSI}
\end{eqnarray}
where the former (latter) ket vector specifies the spin orientation of the localized Ising spin (mobile electrons) from the $i$-th elementary unit.
The paramagnetic state $\rm I$ emerges due to absence of the mobile electrons on the decorating sites, which splits the investigated spin-electron chain into a set of non-interacting Ising spins separated from each other by the empty decorating sites. The alignment of localized Ising spins solely comes  from a presence of the external magnetic field. A more interesting spin arrangement can be found in the phase ${\rm II}$
\begin{eqnarray}
{|{\rm II} \rangle} &=& \prod_{i=1}^N {\Bigl|\frac{1}{2} \Bigr\rangle}_{\sigma_i} \otimes \frac{1}{\sqrt{2}} \left( \hat{a}_{i,\uparrow}^{\dagger} + \hat{b}_{i,\uparrow}^{\dagger} \right) {|0,0\rangle}_i, \nonumber \\
E_{\rm II} &=& -\frac{1}{4} - t - \mu - h,
\label{GSII}
\end{eqnarray}
which is characterized by a single hopping electron on each couple of decorating sites coupled ferromagnetically to its neighbouring Ising spins. Another intriguing spin arrangement can be detected in the phase ${\rm III}$
\begin{eqnarray}
\fl |{\rm III} \rangle &=& \prod_{i=1}^{N/2} {\Bigl|\frac{1}{2} \Bigr\rangle}_{\!\!\sigma_{2i-1}} \!\!\!\!\!\!\!\!
\otimes \! \biggl[\! R_{+} \hat{a}_{2i-1,\uparrow}^{\dagger} \hat{b}_{2i-1,\downarrow}^{\dagger} \!\!-\! R_{-} \hat{a}_{2i-1,\downarrow}^{\dagger} \hat{b}_{2i-1,\uparrow}^{\dagger} \!+\! \frac{2t}{P} \! \left( \hat{a}_{2i-1,\uparrow}^{\dagger} \hat{a}_{2i-1,\downarrow}^{\dagger} \!+\! \hat{b}_{2i-1,\uparrow}^{\dagger} \hat{b}_{2i-1,\downarrow}^{\dagger} \!\right) \!\biggr] {|0,0\rangle}_{2i-1} \nonumber \\ \
\fl &\otimes& {\Bigl|-\frac{1}{2} \Bigr\rangle}_{\!\sigma_{2i}} \!\!\!  \otimes \biggl[- R_{-} \hat{a}_{2i,\uparrow}^{\dagger} \hat{b}_{2i,\downarrow}^{\dagger} + R_{+} \hat{a}_{2i,\downarrow}^{\dagger} \hat{b}_{2i,\uparrow}^{\dagger} 
+ \frac{2t}{P} \left( \hat{a}_{2i,\uparrow}^{\dagger} \hat{a}_{2i,\downarrow}^{\dagger} + \hat{b}_{2i,\uparrow}^{\dagger} \hat{b}_{2i,\downarrow}^{\dagger} \right) \biggr] {|0,0\rangle}_{2i}, \nonumber \\
\fl E_{\rm III} \!\!\!&=&\!\!\! -2 \mu - \frac{P}{2},
\label{GSIII}
\end{eqnarray}
whereas $R_{\pm} = \frac{1}{2} (1 \pm J/P)$. Each pair of the decorating sites is occupied in the phase $\rm III$ by two mobile electrons, which underlie due to a hopping process a quantum superposition of two antiferromagnetic states ($\hat{a}_{i,\uparrow}^{\dagger} \hat{b}_{i,\downarrow}^{\dagger} {|0,0\rangle}_i,\hat{a}_{i,\downarrow}^{\dagger} \hat{b}_{i,\uparrow}^{\dagger} {|0,0\rangle}_i$) and two non-magnetic ionic states ($\hat{a}_{i,\uparrow}^{\dagger} \hat{a}_{i,\downarrow}^{\dagger} {|0,0\rangle}_i,\hat{b}_{i,\uparrow}^{\dagger} \hat{b}_{i,\downarrow}^{\dagger} {|0,0\rangle}_i$). It is quite evident that the hopping process of the mobile electrons gives rise to an antiferromagnetic order of the localized Ising spins and consequently, the phase $\rm III$ has translationally broken symmetry. Owing to the particle-hole symmetry, the ground state of the studied spin-electron chain may also form the phase $\rm IV$ with three mobile electrons per unit cell
\begin{eqnarray}
{|{\rm IV} \rangle} &=& \prod_{i=1}^N {\Bigl|\frac{1}{2} \Bigr\rangle}_{\sigma_i} \otimes \frac{1}{\sqrt{2}} \left( \hat{a}_{i,\uparrow}^{\dagger} \hat{b}_{i,\uparrow}^{\dagger} \hat{b}_{i,\downarrow}^{\dagger} + \hat{a}_{i,\uparrow}^{\dagger} \hat{a}_{i,\downarrow}^{\dagger} \hat{b}_{i,\uparrow}^{\dagger} \right) {|0,0\rangle}_i, \nonumber \\
E_{\rm IV} &=& - 3 \mu - h - t - \frac{1}{4},
\label{GSIV}
\end{eqnarray}
which represents a mirror image of the phase $\rm II$ under assumption that the hopping process of one electron is replaced by a virtual hopping process of one hole. The phase $\rm V$ with fully occupied decorating sites can be characterized by the following state vector
\begin{eqnarray}
{|{\rm V} \rangle} &=&  \Biggl \{ \begin{array}{l}
\displaystyle{\prod_{i=1}^N} {\Bigl|\frac{1}{2} \Bigr\rangle}_{\sigma_i} \otimes \left(\hat{a}_{i,\uparrow}^{\dagger} \hat{a}_{i,\downarrow}^{\dagger} \hat{b}_{i,\uparrow}^{\dagger} \hat{b}_{i,\downarrow}^{\dagger}\right) {|0,0\rangle}_i 
\quad \qquad (\mbox{for} \,\, h > 0), \\
\displaystyle{\prod_{i=1}^N} {\Bigl|\pm \frac{1}{2} \Bigr\rangle}_{\sigma_i} \otimes \left(\hat{a}_{i,\uparrow}^{\dagger} \hat{a}_{i,\downarrow}^{\dagger} \hat{b}_{i,\uparrow}^{\dagger} \hat{b}_{i,\downarrow}^{\dagger}\right) {|0,0\rangle}_i \qquad (\mbox{for} \,\, h = 0), 
\end{array} \Biggr.
\nonumber \\
E_{\rm V} \!\!\!&=&\!\!\! -4 \mu - \frac{h}{2}.
\label{GSV}
\end{eqnarray}
Both decorating sites are fully occupied in the phase $\rm V$ and hence, the mobile electrons transmit zero effective interaction between the localized Ising spins quite similarly as in the phase $\rm I$. The phase ${\rm VI}$ is totally absent in the ground-state phase diagram at small enough hopping terms (c.f. the upper panel in Fig. \ref{fig2} with the central and lower panels), but it may become the relevant ground state on assumption that the hopping term is sufficiently large, the chemical potential is small enough and the external magnetic field is of moderate strength
\begin{eqnarray}
{|{\rm VI} \rangle} &=& \prod_{i=1}^{N} {\Bigl|\frac{1}{2}\Bigr\rangle}_{\sigma_{i}} 
\otimes \frac{1}{2} \left( \hat{a}_{i,\uparrow}^{\dagger} \hat{b}_{i,\downarrow}^{\dagger} - \hat{a}_{i,\downarrow}^{\dagger} \hat{b}_{i,\uparrow}^{\dagger} + \hat{a}_{i,\uparrow}^{\dagger} \hat{a}_{i,\downarrow}^{\dagger} + \hat{b}_{i,\uparrow}^{\dagger} \hat{b}_{i,\downarrow}^{\dagger} \right) {|0,0\rangle}_{i} \nonumber \\
E_{\rm VI} &=& -2 \mu - \frac{h}{2} - 2 t.
\label{GSVI}
\end{eqnarray}
The phase $\rm VI$ is very similar to the phase $\rm III$ except that all localized Ising spins are fully polarized into the magnetic field and consequently, the mobile electrons underlie a symmetric quantum superposition of two antiferromagnetic and two ionic states with four equal probability amplitudes. Finally, the phase $\rm VII$  becomes the ground state of the investigated spin-electron chain at sufficiently high magnetic fields 
\begin{eqnarray}
{|{\rm VII} \rangle} &=& \prod_{i=1}^N {\Bigl|\frac{1}{2}\Bigr\rangle}_{\sigma_{i}}  \otimes \left( \hat{a}_{i,\uparrow}^{\dagger} \hat{b}_{i,\uparrow}^{\dagger} \right) {|0,0\rangle}_i, \nonumber \\
E_{\rm VII} &=& -2 \mu - \frac{3 h}{2} - \frac{1}{2}.
\label{GSVII}
\end{eqnarray}
The phase $\rm VII$ involves two mobile electrons per each couple of the decorating sites, which align their spins into the external magnetic field quite similarly as the localized Ising spins. For completeness, analytical expressions for the first-order phase boundaries between the respective ground states are listed below 
\begin{eqnarray}
{\rm I} - {\rm II}&:& \qquad \mu = - \frac{1}{4} - t - \frac{h}{2} \nonumber \\
{\rm II} - {\rm III}&:& \qquad \mu = \frac{1}{4} + t - \frac{P}{2} + h \nonumber \\
{\rm III} - {\rm IV}&:& \qquad \mu = - \frac{1}{4} - t + \frac{P}{2} - h \nonumber \\
{\rm IV} - {\rm V}&:& \qquad \mu = \frac{1}{4} + t + \frac{h}{2} \nonumber \\
{\rm II} - {\rm VI}&:& \qquad \mu = \frac{1}{4} - t + \frac{h}{2} \nonumber \\
{\rm II} - {\rm VII}&:& \qquad \mu = - \frac{1}{4} + t - \frac{h}{2} \nonumber \\
{\rm IV} - {\rm VI}&:& \qquad \mu = - \frac{1}{4} + t - \frac{h}{2} \nonumber \\
{\rm IV} - {\rm VII}&:& \qquad \mu = \frac{1}{4} - t + \frac{h}{2}.
\label{Boundaries}
\end{eqnarray}

\subsection{Magnetization process}

\begin{figure}
\begin{center}
\includegraphics[width=10cm]{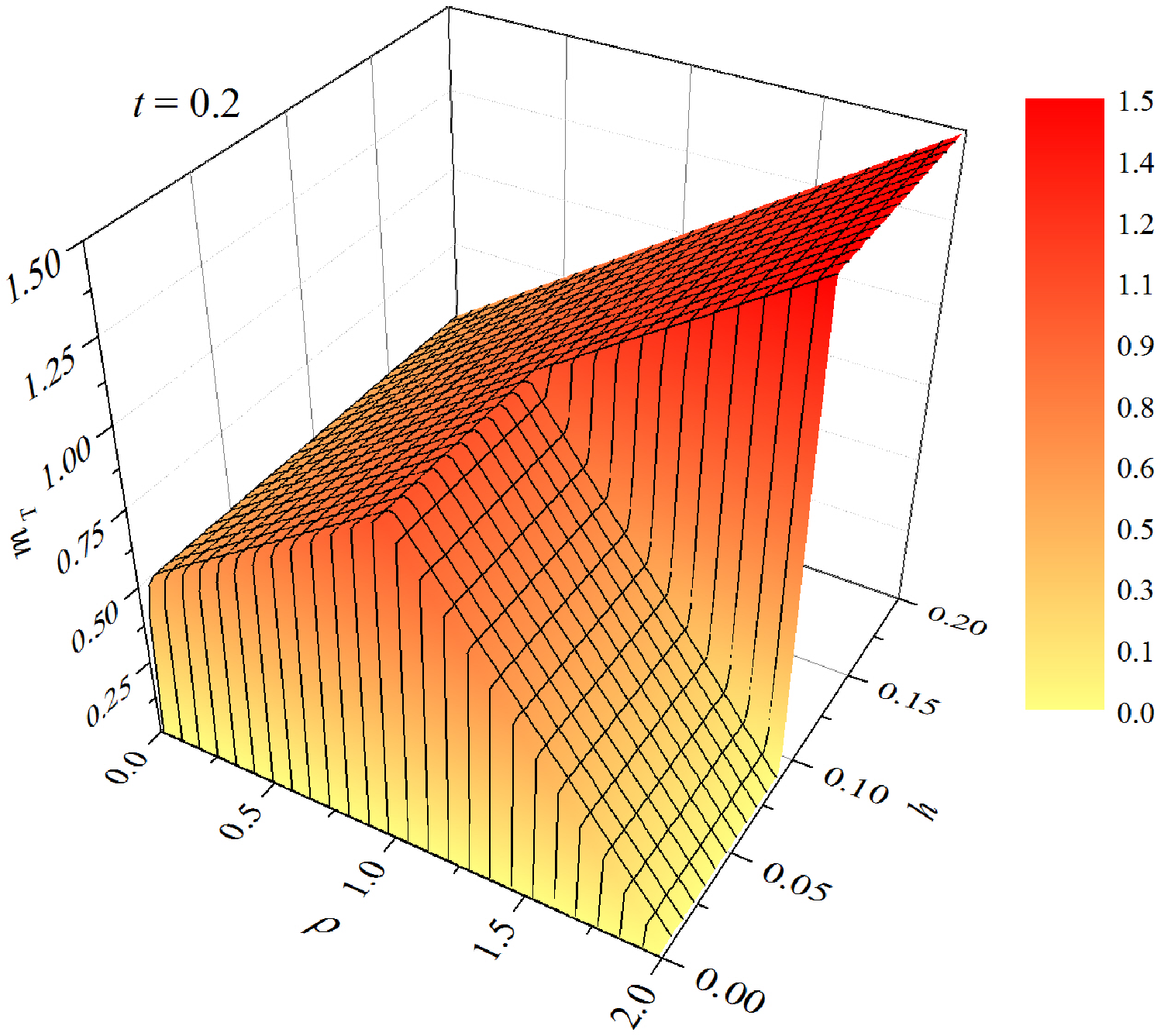}
\vspace{-0.2cm}
\includegraphics[width=10cm]{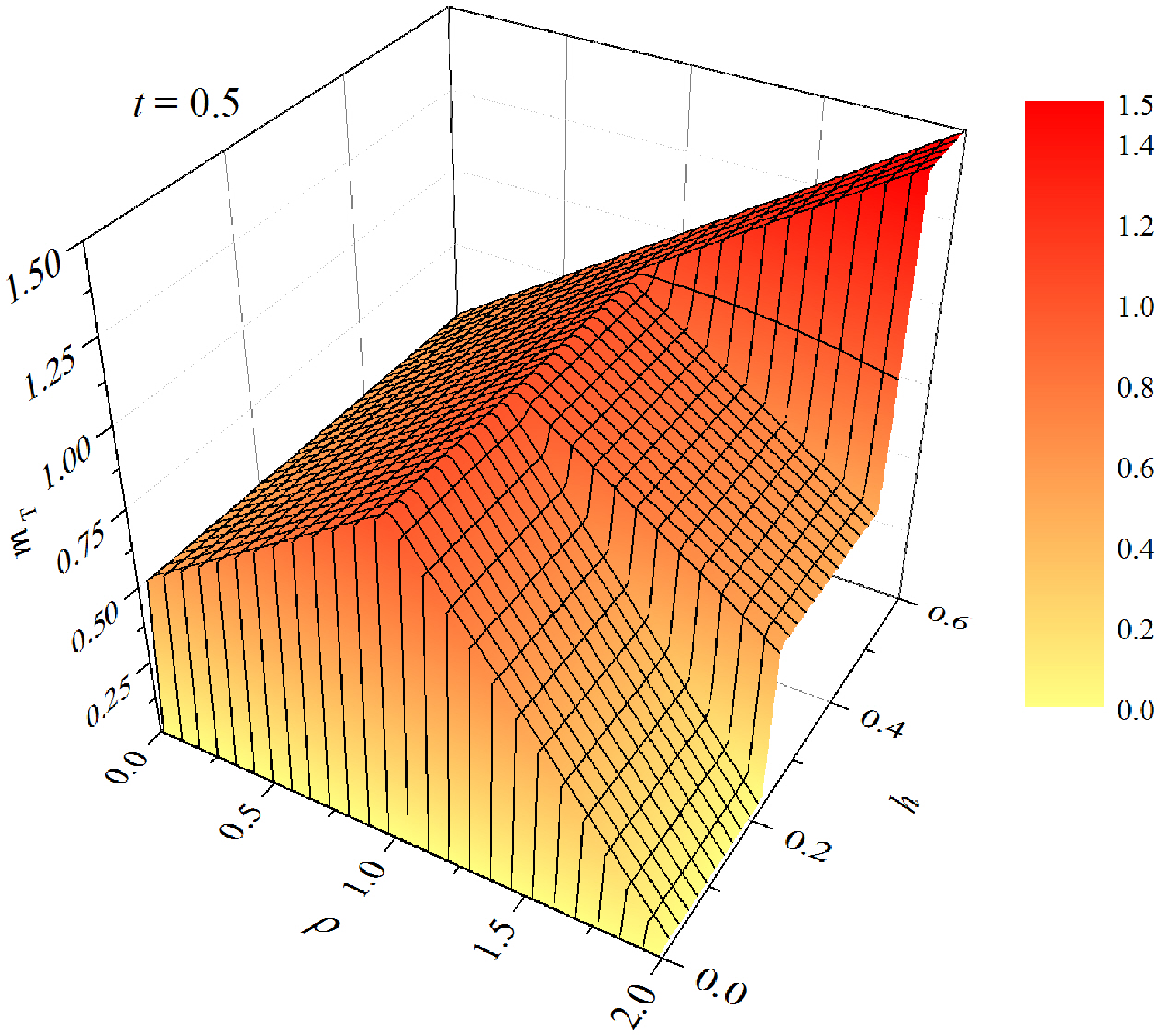}
\vspace{-0.2cm}
\includegraphics[width=10cm]{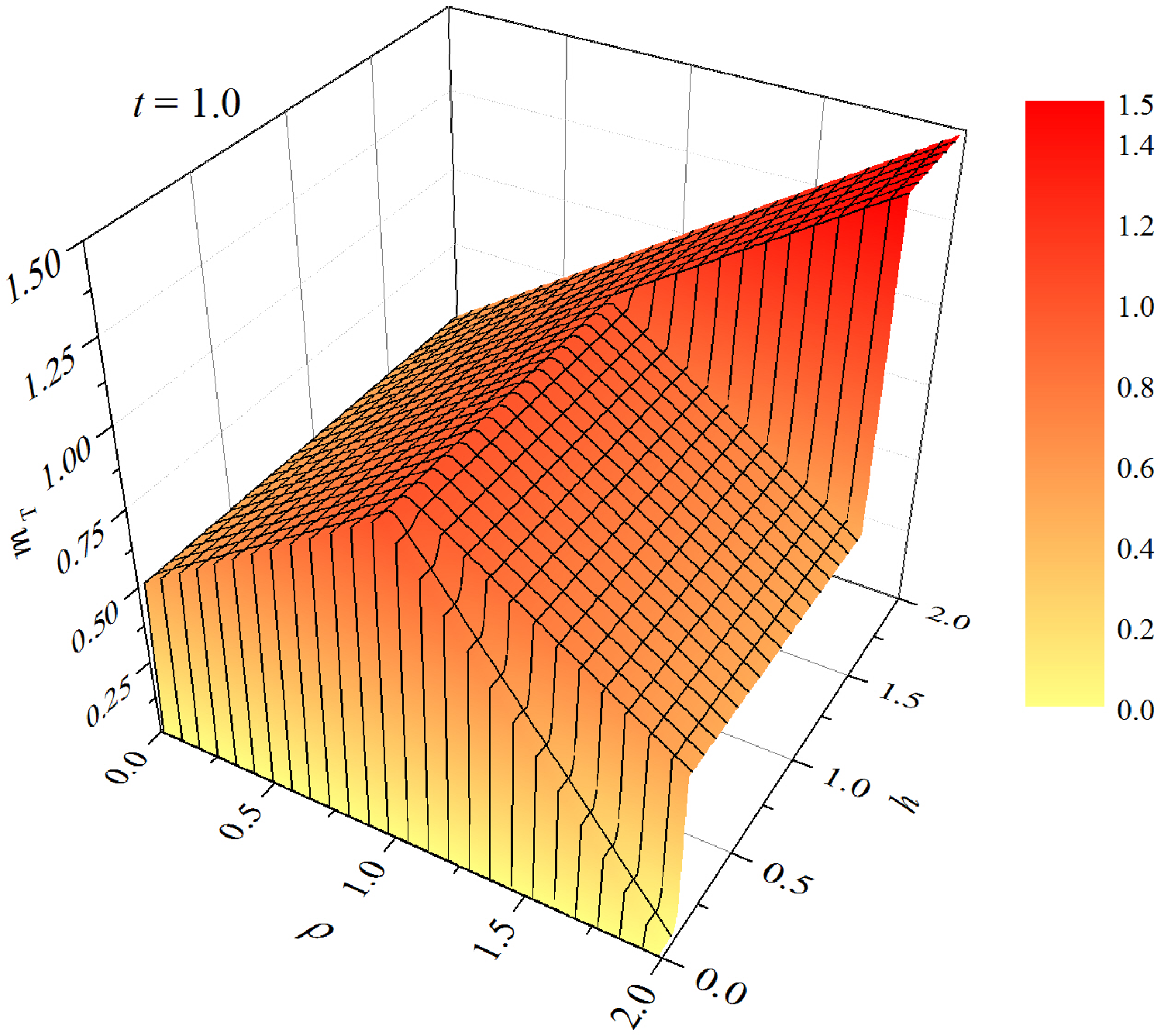}
\end{center}
\vspace{-0.6cm}
\caption{\small (Color online) 3D surface plot of the total magnetization $m_T$ per unit cell as a function of the electron density $\rho$ and the magnetic field $h$ at low enough temperature $T=0.01$
for three particular values of the hopping term $t=0.2, 0.5$ and $1.0$.}
\label{Magnetization}
\end{figure}

Next, let us proceed to a discussion of the low-temperature magnetization process of the investigated spin-electron chain. For this purpose, we have plotted in Fig.~{\ref{Magnetization}} the total magnetization $m_T$ per elementary unit against the electron density $\rho$ and the magnetic field $h$ for three different values of the hopping term $t = 0.2, 0.5$ and $1.0$ at relatively low temperature $T=0.01$. At sufficiently small electron densities $0 \leq \rho \leq 1.0$ the total magnetization starts from zero and abruptly reaches its saturated value with increasing of the magnetic field, because all localized Ising spins as well as mobile electrons are aligned into the magnetic-field direction. Contrary to this, the magnetization scenario is much more complex at higher values of the doping parameter $\rho$, where the shape of magnetization curve basically depends on a relative strength of the kinetic term $t$. 

For sufficiently low values of the hopping term (e.g. $t = 0.2$ in the upper panel of Fig. \ref{Magnetization}) one generally observes just one intermediate plateau before the total magnetization reaches the saturated value upon strengthening of the external magnetic field. To get an insight into the ground state corresponding to the particular magnetization plateau, let us take a closer look at two limiting cases with the electron density $\rho = 1.0$ and $2.0$ corresponding to a quarter and a half filling of the decorating sites. The observed plateau at a quarter filling ($\rho = 1.0$) is consistent with the ferromagnetic phase $\rm II$ involving a single hopping electron per elementary unit, while the zero magnetization plateau at a half filling ($\rho = 2.0$) appears due to the quantum antiferromagnetic phase $\rm III$ with two hopping electrons per unit cell. It could be thus concluded that a doping with the mobile electrons gradually increases the number of the effective antiferromagnetic bonds occupied by two hopping electrons at the expense of the effective ferromagnetic bonds mediated by a single hopping electron. In this way, the height of doping-dependent magnetization plateau can be tuned continuously.  

Even more striking low-temperature magnetization curves with two intermediate plateaus can be detected at higher values of the hopping term (e.g. $t = 0.5$ and $1.0$ in the central and lower panel of Fig. \ref{Magnetization}). The first magnetization plateau at lower magnetic fields is of the same origin as described previously, i.e. the doping with the mobile electrons causes a gradual onset of the antiferromagnetic spin order of the localized Ising spins due to the effective antiferromagnetic coupling mediated by the bonds occupied with two hopping electrons with an opposite spins. On the other hand, the additional intermediate plateau observed at moderate magnetic fields emerges owing to a mutual competition between the magnetic field and the quantum-mechanical hopping process. The magnetic field of the moderate strength is strong enough to polarize all localized Ising spins towards the magnetic-field direction, but simultaneously it does not suffice to break the antiferromagnetic alignment on the bonds occupied with two hopping electrons of opposite spins. As a matter of fact, the one-third magnetization plateau existing at a half filling of the decorating sites ($\rho = 2.0$) is consistent with the ferrimagnetic ground state $\rm VI$ given by Eq. (\ref{GSVI}). In accordance with this statement, the second magnetization plateau extends over a wider range of the magnetic fields upon strengthening of the hopping term $t$, which makes an opposite spin alignment of two mobile electrons more favourable. To summarize this part, we have verified that the doping of decorating sites of the spin-1/2 Ising chain with the mobile electrons allows one to control the height of the fractional plateaus in the low-temperature magnetization curve even if the mobile electrons are delocalized only over the pairs of decorating sites.

\subsection{Magnetocaloric effect}

\begin{figure}
\begin{center}
\includegraphics[width=10.0cm]{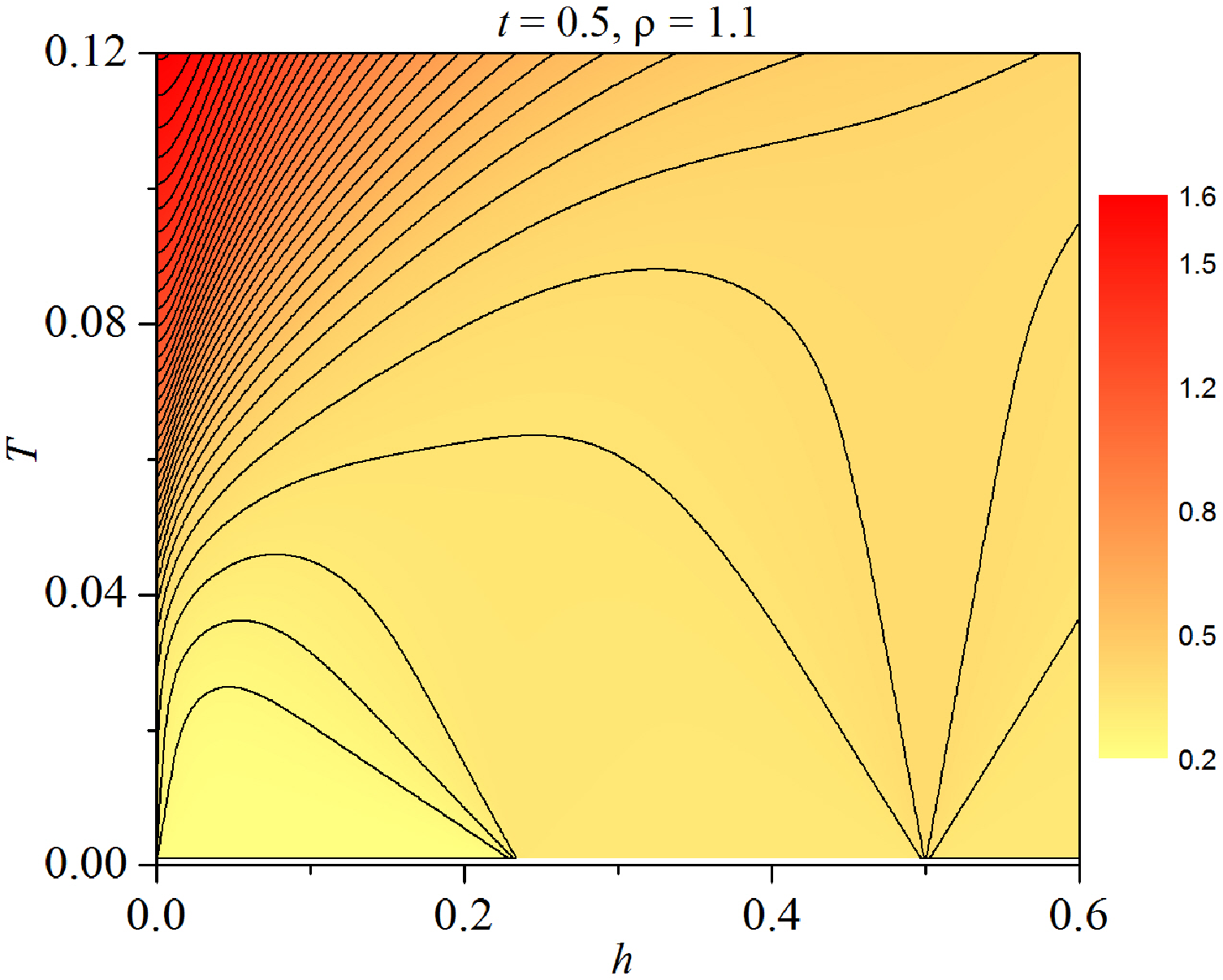}
\vspace{-0.2cm}
\includegraphics[width=10.0cm]{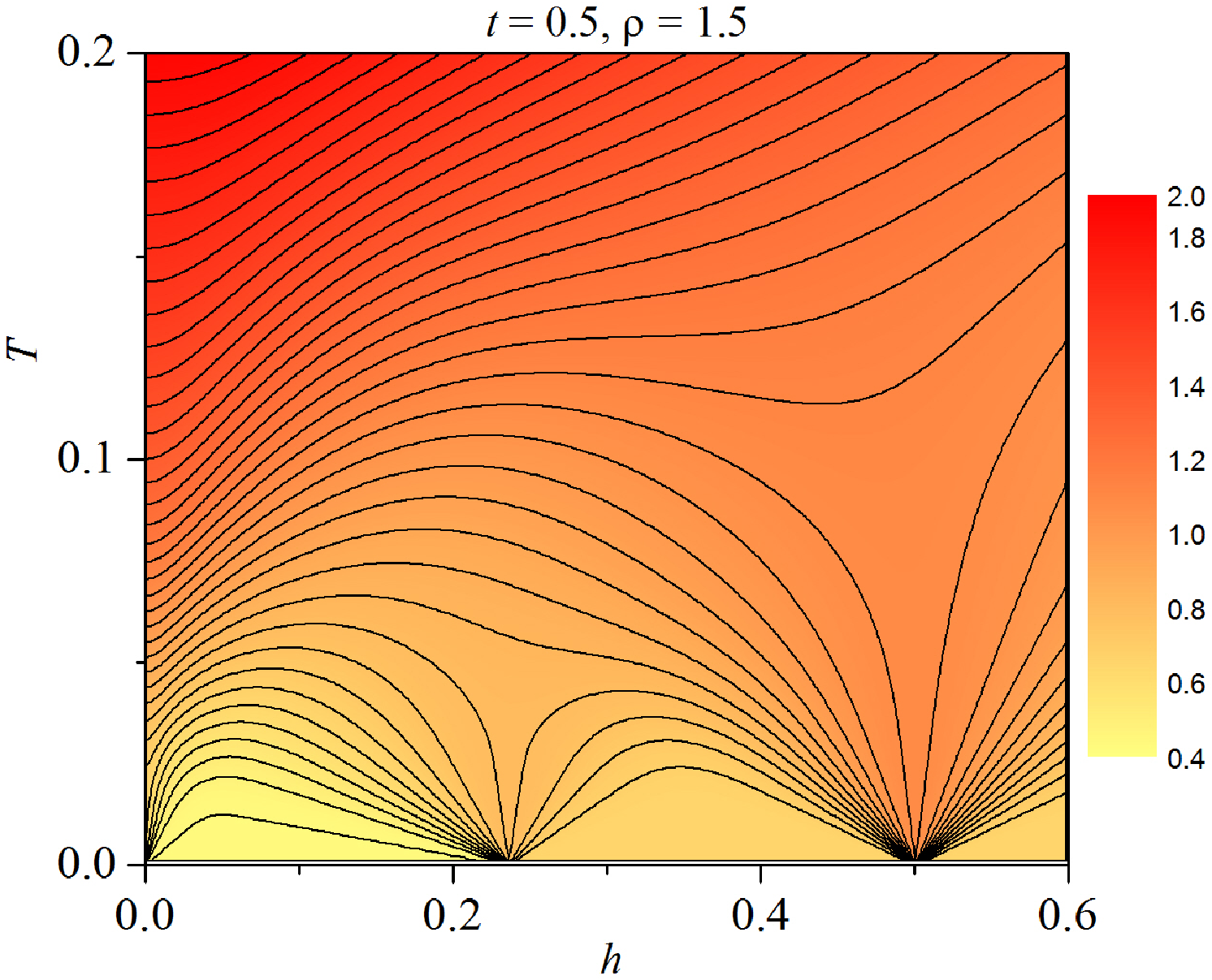}
\vspace{-0.2cm}
\includegraphics[width=10.0cm]{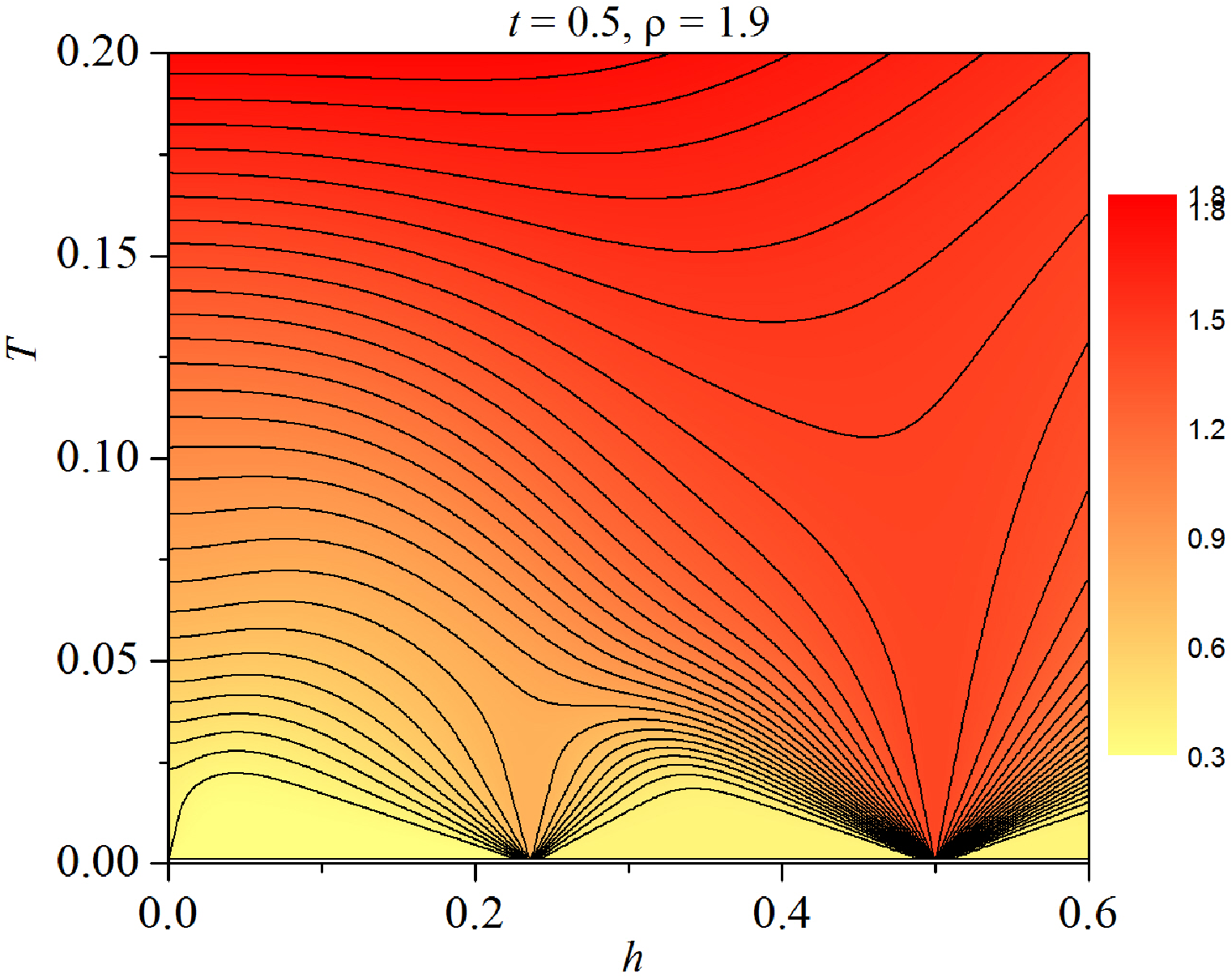}
\end{center}
\vspace{-0.6cm}
\caption{\small (Color online) The adiabatic (isentropic) changes of temperature as a function of magnetic field for the fixed value of the hopping term $t=0.5$ and three different values of the electron density $\rho=1.1, 1.5$ and $1.9$.}
\label{entropy}
\end{figure}

Last but not least, let us investigate in detail an adiabatic change of the temperature with the external magnetic field as the most significant characteristics of the magnetocaloric effect. Fig. {\ref{entropy}} displays typical isentropic dependences of temperature on the magnetic field for the hopping term $t=0.5$ and three different electron densities $\rho=1.1, 1.5$ and $1.9$. As one can see, the most abrupt drop (rise) of the temperature upon decreasing of the magnetic field can be found for all three electron densities above (below) the critical fields $h = 0.0, 0.23$ and $0.5$, which perfectly coincide with the field-driven magnetization jumps between the individual magnetization plateaus. A vigorous drop of the temperature upon the adiabatic demagnetization observable slightly above zero magnetic field implies an intriguing refrigeration potential of the studied spin-electron system, which is capable of reaching ultra-low temperatures down to the absolute zero temperature. The main difference between the displayed cases with three different electron densities lies in the temperature range, over which the fast cooling effect can be achieved by the adiabatic demagnetization. It turns out that the annealed bond disorder generally enhances the refrigeration capability in comparison to the special cases without the annealed bond disorder.   

\section{Concluding remarks}
\label{sec:conclusions}

In the present work, the coupled spin-electron chain consisting of localized Ising spins and mobile electrons has been exactly solved in an external magnetic field within the transfer-matrix method. It has been shown that the ground-state phase diagram involves seven different phases, which differ by the number of mobile electrons per unit cell and respective spin arrangements. In particular, we have rigorously studied how the low-temperature magnetization curves depend on the electron doping and magnetic field. It has been demonstrated that the magnetization curves may involve one or two intermediate magnetization plateaus, whose height may be continuously tuned by the electron doping. Besides, we have exactly examined the adiabatic change of temperature upon varying of the external magnetic field. The enhanced magnetocaloric effect has been detected close to all field-driven phase transitions, whereas the annealed bond disorder turns out to enhance the fast cooling down to the absolute zero temperature. Although we are currently not aware of any magnetic material that would afford an experimental realization of the considered spin-electron chain, the targeted design achieved by a chemical reduction of binuclear dicopper core of the polymeric coordination compound MnCu$_2$(bapo)(H$_2$O)$_4$.2H$_2$O [bapo=N,N'-bis(oxamato-1,3-propylene)oxamide] \cite{peiy86,geor89} could represent an intriguing possibility for an experimental testing of doping-dependent magnetization plateaus.

\section*{Acknowledgments}
This work was financially supported by the grant of The Ministry of Education, Science, Research and Sport of the Slovak Republic under the contract No. VEGA 1/0043/16 and by the grant of the Slovak Research and Development Agency provided under Contract No. APVV-0097-12.

\end{document}